\documentclass[aps,pre,twocolumn,groupedaddress,showpacs,amsmath,amssymb]{revtex4}

\usepackage{graphicx}
\usepackage{dcolumn}
\begin{document}

\title{Allelomimesis as universal clustering mechanism\\ 
for complex adaptive systems}
\author{Dranreb Earl Juanico, Christopher Monterola, Caesar Saloma}
\homepage[Fax:]{+632 9205474}
\email{csaloma@nip.upd.edu.ph}
\affiliation{National Institute of Physics, University of the Philippines\\
Diliman, Quezon City, Philippines 1101}

\date{\today}

\begin{abstract}
Animal and human clusters are complex adaptive systems and many are organized
in cluster sizes $s$ that obey the frequency-distribution $D\left(s\right)\propto s^{-\tau}$.
Exponent $\tau$ describes the relative abundance of the cluster sizes in a given
system. Data analyses have revealed that real-world clusters exhibit a broad spectrum
of $\tau$-values, $0.7\textrm{ (tuna fish schools)}\leq\tau\leq 2.95\textrm{ (galaxies)}$.
We show that allelomimesis is a fundamental mechanism for adaptation that accurately
explains why a broad spectrum of $\tau$-values is observed in animate, human and 
inanimate cluster systems. Previous mathematical models could not account
for the phenomenon. They are hampered by details and apply only to
specific systems such as cities, business firms or gene family sizes. Allelomimesis is the tendency
of an individual to imitate the actions of its neighbors and two cluster systems 
yield different $\tau$ values if their component agents display different allelomimetic
tendencies. We demonstrate that allelomimetic adaptation are of three
general types: blind copying, information-use copying, and non-copying. Allelomimetic
adaptation also points to the existence of a stable cluster size consisting of three 
interacting individuals.
\end{abstract}


\maketitle

\section{Introduction}
Huge amounts of data have been collected and analyzed by researchers in various fields
of the natural and social sciences concerning the clustering behavior of animals
(fish schools, buffalo herds, etc). In the real world, a number of different cluster types
often exist and share common habitat and an accurate understanding of cluster formation
among diverse animate and inanimate systems is of great value in wildlife preservation,
environmental management, urban planning, economics, genetics and even politics.

Animal and human clusters are complex systems with adaptive agents. Many exist in cluster
sizes $s$ that obey the frequency distribution $D\left(s\right)\propto s^{-\tau}$. Exponent $\tau$
determines the relative abundance of the cluster sizes -- a small $\tau(\approx 0)$ implies
equal abundance of large and small clusters in a given system. Power-law distributions indicate
the role of self-organization during cluster formation and the presence of scale-free interaction
dynamics that holds over several scales of the agent population and size of interaction 
space \cite{bak96,stanley96}.

Scale-free clusters have been observed with gene families, colloids, fish schools, slum areas,
city populations, business firms, and galaxies to name a few. Table \ref{tab::real-world} lists forty-five different 
real-world cluster systems with their measured $\tau$ values ranging from $\tau=0.7$ (tuna fish
schools) \cite{bonabeau95,bonabeau99} to $\tau=2.95$ (galaxy clusters) \cite{jarrett}.

To our knowledge, no mathematical model can generate size-frequency distributions that cover the
entire range of $\tau$-values in Table \ref{tab::real-world}. Available models are effective only at describing 
particular systems such as cities \cite{stanley96,johansson97}, firms \cite{axtell01}, or gene
families \cite{huynen98}. Current models utilize many interaction details that limit the range
of their applicability \cite{may04}.

Here, we show that allelomimesis is a fundamental mechanism for adaptation that can accurately 
explain why real-world cluster systems exhibit a broad spectrum of $\tau$-values. Allelomimesis
is the act of copying one's kindred neighbors \cite{parrish99,juanico03}. We demonstrate that 
differences in $\tau$-values are caused from variations in allelomimetic behavior (described by
a single parameter $\alpha$ where $0\leq\alpha\leq 1$) of the agent phenotypes from one cluster
system to another. We derive a nonlinear relation between $\tau$ and $\alpha$ that rationalizes
the distribution of $\tau$ values in animal, human, and inanimate cluster systems. For strongly-
allelomimetic agents ($\alpha\approx 1$), clustering by allelomimesis predicts a stable cluster
size at $s=3$ which has been observed previously in marmots \cite{grimm03} and killer whales
\cite{baird96}.

Our clustering model unifies previous mathematical models of cluster formation towards a common
starting point. Allelomimesis is expressed in various (higher-order) forms in previous 
self-aggregation models. In animal aggregation models, it is manifested as biosocial attraction
\cite{bonabeau99} or as conspecific copying \cite{wagner03}. For example, herding which has been
observed in panicking mice escaping from a two-door enclosure is a striking example of nearest-
neighbor copying \cite{saloma03}. In the percolation model of urban growth, allelomimesis is 
implicit in the concept of correlation \cite{makse95}. The tendency of employees to associate
with those that belong to the same income bracket in Axtell's model of firms \cite{axtell01} can
also be construed as another expression of allelomimesis.

Copying is natural among social groups and is an evolutionary mechanism in human societies \cite{higgs00}.
In a community of strongly allelomimetic individuals, it is natural to expect that full cooperation
is achieved quickly without the threat of punishment \cite{milinski02}. In gene families, cluster
formation is explained as an intricate birth-death process \cite{huynen98}. In olivine crystal 
sizes, it is driven by complicated tectonic processes \cite{armienti02}.

Allelomimetic interaction between agents could be described with few simple local rules \cite{juanico03}.
A single measure $\alpha$ is sufficient to vary the exponent $\tau$ over a wide range of values. Agents
search for neighbors (kindreds) and those that are strongly-allelomimetic ($\alpha\approx 1$) are likely
to copy their neighbors leading to the formation of relatively large clusters such as those observed in 
fish schools (see Table \ref{tab::real-world}). On the other hand, relatively large clusters are highly unlikely in
gene families, colloids, and galaxies which are systems with components that are incapable of copying
each other ($\alpha\approx 0$).

Of particular interest are human cluster systems such as slums of informal settlers, cities, and
business firms. Table \ref{tab::real-world} reveals that such systems occupy a narrow band of the $\tau$-spectrum
($1.4\leq\tau\leq 2.16$). Unlike other animals, humans are rational and capable of deciding on their
own based upon a set of competing factors that promote individual self-interest on the one hand and
collective benefit on the other. We determine the possible classes of allelomimetic interactions by
establishing a quantitative relation between $\alpha$ and $\tau$.

Our presentation proceeds as follows: In Sec.\ \ref{sec::methodology} we describe briefly our 
mathematical model for clustering by allelomimesis while in Sec.\ \ref{sec::results} we compare the
predictions of the model with those observed in real-world clusters such as those listed in Table \ref{tab::real-world}.
We end our presentation by discussing the results of our comparison.

\section{Methodology}
\label{sec::methodology}
 \begin{table*}
 \caption{Measured $\tau$-values of real-world cluster systems with power-law
 $D\left(s\right)$ plots. Also shown are the corresponding values of the allelomimesis measure
 $\alpha$.\label{tab::real-world}}
 \begin{ruledtabular}
 \begin{tabular}{lrr}
 REAL-WORLD CLUSTER SYSTEM & $\tau$ & $\alpha$\\\hline
 Tuna near fish-aggregating device \cite{bonabeau95,bonabeau99}& $0.70$ & $0.9895$ \\\hline
 Clupeid fish \emph{Sardinella maderensis} \& \emph{S. aurea} \cite{bonabeau95,bonabeau99}&
 					$0.95$ & $0.9895$\\\hline
 Alpine marmot \emph{Marmota marmota} \cite{grimm03} & $1.08\pm0.25$ & $0.9895$ \\\hline
 African bufallo \emph{Syncerus cafer} \cite{bonabeau99} & $1.15$ & $0.9894$ \\\hline
 Wasps \emph{Ropalidia fasciata} \cite{ito02} & $1.19\pm0.008$ & $0.9893$ \\\hline
 Tephritid flies \cite{sjoberg00} & $1.37$ & $0.9888$ \\\hline
 Free-swimming tuna with three species mixed \cite{bonabeau95,bonabeau99} & $1.49$ & $0.9877$ \\\hline
 Spatial sizes of forest fires \cite{malamud98} & $1.5$ & $0.9877$ \\\hline
 Offshore-spotted dolphin \emph{Stenella attenuata} \cite{perkins97} & $1.79\pm0.05$ & $0.9590$ \\\hline
 Three species of African baboons \cite{altmann70} & $2.01\pm0.08$ & $0.7841$ \\\hline
 West Indian manatee \emph{Trichecus manatus} \cite{kadel91} & $2.19\pm0.07$ & $0.4009$ \\\hline
 Nomadic Serengeti cheetah \emph{Acinonyx jubatus} \cite{schaller72} & $2.49\pm0.35$ & $0.0378$ \\\hline
 Nomadic Serengeti lion \emph{Panthera leo} \cite{schaller72} & $2.49\pm0.08$ & $0.0379$ \\\hline
 Mathare valley squatter settlements, Kenya \cite{sobreira01} & $1.40\pm0.20$ & $0.9888$ \\\hline
 Recife squatter settlements, Brazil \cite{sobreira01} & $1.60\pm0.20$ & $0.9844$ \\\hline
 French manufacturing firms in 1962 \cite{vanark96} & $1.84\pm0.08$ & $0.9411$ \\\hline
 Japanese manufacturing firms in 1975 \cite{vanark96} & $1.90\pm0.04$ & $0.9067$ \\\hline
 Urban agglomerations, India \cite{brinkhoff} & $1.93\pm0.002$ & $0.87$ \\\hline
 Towns surrounding London in 1981 \cite{stanley96,makse95} & $1.96$ & $0.8514$ \\\hline
 Towns surrounding Berlin in 1981 \cite{stanley96,makse95} & $1.98$ & $0.8270$ \\\hline
 Swedish firms in 1993 \cite{johansson97} & $1.98\pm0.08$ & $0.8270$ \\\hline
 Urban areas in Great Britain in 1981 \& 1991 \cite{makse95} & $2.03$ & $0.7512$ \\\hline
 City populations in Brazil in 1991 \& 1993 \cite{brinkhoff} & $2.04\pm0.06$ & $0.7329$ \\\hline
 Urban agglomerations in Russia in 1994 \cite{brinkhoff} & $2.04\pm0.06$ & $0.7285$ \\\hline
 Urban agglomerations in U.S.A. in 1994 \cite{brinkhoff} & $2.04\pm0.07$ & $0.7250$ \\\hline
 Urban agglomerations in France in 1982 \& 1990 \cite{brinkhoff} & $2.05\pm0.00$ & $0.7049$ \\\hline
 U.S. firms in 1997 \cite{axtell01} & $2.06\pm0.05$ & $0.6974$ \\\hline
 City populations in Mexico in 1990 \cite{brinkhoff} & $2.07\pm0.00$ & $0.6814$ \\\hline
 City populations in China in 1990 \cite{brinkhoff} & $2.11\pm0.00$ & $0.5970$ \\\hline
 World's most populous cities in 2002 \cite{brinkhoff} & $2.11\pm0.08$ & $0.5959$ \\\hline
 British business/manufacturing firms in 1955 \cite{simon58} & $2.11$ & $0.5881$ \\\hline
 City populations in Germany in 1994 \cite{brinkhoff} & $2.15\pm0.20$ & $0.5033$ \\\hline
 City populations in Japan in 1994 \cite{brinkhoff} & $2.16\pm0.09$ & $0.4692$ \\\hline
 Gene family sizes of \emph{M. pneumoniae} \cite{huynen98} & $2.69$ & $0.0056$ \\\hline
 Gene family sizes of \emph{S. cerevisiae} \cite{huynen98} & $2.81$ & $0.0018$ \\\hline
 Gene family sizes of \emph{E. coli} \cite{huynen98} & $2.84$ & $0.0013$ \\\hline
 Gene family sizes of \emph{Synechocystis sp.} \cite{huynen98} & $3.17$ & $0.0001$ \\\hline
 Gene family sizes of \emph{H. influenzae} \cite{huynen98} & $3.27$ & $0.0000$ \\\hline
 Gene family sizes of \emph{M. janaschii} \cite{huynen98} & $3.62$ & $0.0000$ \\\hline
 Gene family sizes of \emph{Vaccinia} virus \cite{huynen98} & $3.80$ & $0.0000$ \\\hline
 Gene family sizes of \emph{M. genitalum} \cite{huynen98} & $4.02$ & $0.0000$ \\\hline
 Gene family sizes of \emph{T4 bacteriophage} \cite{huynen98} & $4.61$ & $0.0000$ \\\hline
 Olivine crystal sizes of GP45 - granoblastic \cite{armienti02} & $2.83\pm0.16$ & $0.0015$ \\\hline
 Olivine crystal sizes of GP30 - coarse \cite{armienti02} & $3.03\pm0.18$ & $0.0002$ \\\hline
 Olivine crystal sizes of LANZ3 - porphyroclastic \cite{armienti02} & $3.81\pm0.41$ & $0.0000$ \\\hline
 Database of 8,000 galaxy clusters \cite{jarrett} & $2.95\pm0.36$ & $0.0005$ \\\hline
 \end{tabular}
 \end{ruledtabular}
 \end{table*}

\subsection{Agent-based model of allelomimetic interaction}
An $N\times N$ lattice with \emph{free} boundaries \cite{domb73} is utilized as a platform for our 
agent-based model of cluster formation. Initially at time step $q_0 = 0$, the lattice cells are empty
and for every subsequent time step $q$, an agent is injected into a randomly-selected vacant cell. An
agent at cell location $\left(x,y\right)$ is assigned a state $\Psi_m\left(x,y\right)$ that is taken
from a set of possible states $\left\{\Psi_m\right\} = \left\{1,2,\ldots, M\right\}$. State 
$\Psi\left(x,y\right)$ represents a particular trait, preference, action or any other social attribute
that characterize an agent at any given time. For example, $\left\{\Psi_m\right\}$ could be different
species of tuna swimming within the same field of observation \cite{bonabeau95} or the different types
of behavior in groups of lions (e.g., hunting, sleeping, yawning, etc.) as described by Schaller 
\cite{schaller72}. An empty cell is assigned the value of $\Psi=0$.

An agent at $\left(x,y\right)$ searches for other agents of similar state by occupying the next available
cell of its Moore neighborhood which consists of the agent's eight nearest cells at locations, 
$\left\{\left(x+j,y+k\right)\right\}$, where indices $i,j = -1,0,1$. The location 
$\left(x+0,y+0\right)= \left(x,y\right)$ represents the current (default) location of the agent. In 
deciding to occupy a neighboring cell $\left(x+j,y+k\right)$ in the next time step $q+1$, the agent 
evaluates the viability of its current position $\left(x,y\right)$ with those of its neighboring cells 
using the discrete potential function $\Phi$,
\begin{widetext}
\begin{equation}
\label{eq::potential}
\Phi\left(x+j,y+k\right) = 100\left\{1-\delta\left[\Psi\left(x+j,y+k\right)\right]\right\}-
\delta\left[\Psi\left(x+j,y+k\right)\right]\sum_{u,v}
\left\{2\delta\left[\Psi\left(x,y\right)-\Psi\left(x+j+u,y+k+v\right)\right]\right\} 
\end{equation}
\end{widetext}

\noindent where $\delta\left(w\right)$ is the Dirac delta function which is non-zero and equal to unity
only when $w=0$. The summation in Eq.\ \ref{eq::potential} is taken from $u=-1, v=-1$ to $u=1, v=1$ and
the possible values for $u$ and $v$ are, $-1,0,1$.

The potential barrier $\Phi\left(x+j,y+k\right)$ is highest at $100$ when $\Psi\left(x+j,y+k\right)\neq 0$
which happens when the cell at $\left(x+j,y+k\right)$ is unavailable for occupation. On the other hand,
$\Phi\left(x+j,y+k\right)$ becomes much less than 100 when $\Psi\left(x+j,y+k\right)=0$ which is the case
when the cell at location $\left(x+j,y+k\right)$ is vacant.

The set of cells $\left\{\left(x+j+u,y+k+v\right)\right\}$ represents the Moore neighborhood of the agent's
neighboring cell $\left(x+j,y+k\right)$. The term, 
$\delta\left[\Psi\left(x,y\right)-\Psi\left(x+j+u,y+k+v\right)\right]$ is a comparison between $\Psi\left(x,y\right)$
with those in the Moore neighborhood of the vacant cell at $\left(x+j,y+k\right)$. It is equal to unity when
$\Psi\left(x,y\right)=\Psi\left(x+j+u,y+k+v\right)$. More details of the model may be found elsewhere
\cite{juanico03}.

Eq.\ \ref{eq::potential} is applied to every cell in the Moore neighborhood and the agent occupies the vacant
cell that yields the lowest (most negative) value for $\Phi$, expressed in the following minimum condition:
\begin{equation}
\label{eq::minimum}
\Phi_{min} = \min_{j,k}\Phi\left(x+j,y+k\right)
\end{equation}

\begin{figure} 
 \includegraphics[width=8.6cm]{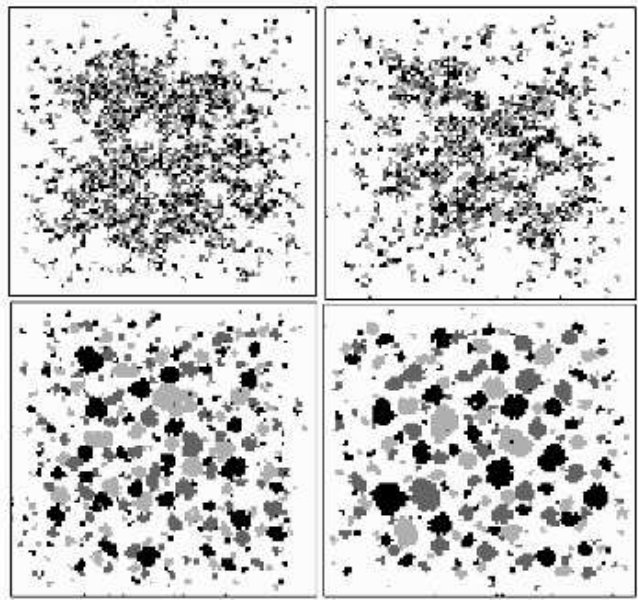}
 \caption{Typical configurations of agents with $\alpha=0$ (first row, first column), 
 0.5 (first row, second column), 0.99 (second row, first column) and 0.995 (second row, second column)
 after $q_T = 600,000$ iterations ($N=100$). Different gray-levels represent three ($M=3$)
 possible states of an agent.
 \label{fig::cluster-config}}
 \end{figure}

\noindent If more than one cell satisfies Eq.\ \ref{eq::minimum}, then the agent randomly selects among these
cells. Successive application of the above-mentioned mechanism for a sufficiently long period of time gives
rise to clustered configurations such as those in Fig.\ \ref{fig::cluster-config}. A cluster is defined 
as a contiguous group of agents of the same state $\Psi$ that are connected via neighboring cells.

The tendency to copy its neighbors can vary from one agent phenotype to another. We use a single parameter
$\alpha$ to describe different levels of allelomimetic behavior where, 
$0\textrm{ (non-allelomimetic)}\leq\alpha\leq 1\textrm{ (blind copying)}$. The state of a completely allelomimetic
agent ($\alpha=1$) always depends on the condition of its Moore neighborhood while that of a non-allelomimetic
agent is oblivious of the states of its Moore neighborhood. For computational simplicity, we assume that all
the agents in the lattice have the same $\alpha$ value. Agent segregation and clustering are attributed
directly to local behavior of the individual agents instead of a globally-defined probability of 
segregation and clustering \cite{bonabeau95}.

Let $\zeta$ be a random variable taken from a uniform distribution between 0 and 1. If $\zeta>\alpha$, then
an agent randomly selects a state from the set $\left\{\Psi_m\right\}$. On the other hand, if $\zeta\leq\alpha$,
then the agent's state is set by its Moore neighborhood such that the state with the highest occurence 
within the neighborhood is the one copied by the agent. If more than one state satisfies this condition,
then the agent selects randomly from among these states. 

To avoid completely filling the lattice through the constant addition of agents, we also include a constant
probability of death for each agent. The death probability is held sufficiently low (1 in 10,000 for every
time step) to maintain a relatively high lattice population density $\rho$. When an agent dies, its cell
location becomes empty in the next time step. The agent population density $\rho$ in the lattice remains
increasing with $q$ and agents are more likely to stay in its original position after a sufficiently long
period of time where number of available vacant cells in its Moore neighborhood becomes smaller.

Each simulation is run for a time-step duration of $q_T = 600,000$ which is sufficient to allow the 
agent population density to reach a steady-state value, $\rho\left(q_T\right)\approx 0.5$. To incrase
the probability of conspecific agents meeting and coalescing within a reasonably short period of 
computational time, we choose $M=3$, i.e., $\left\{\Psi_m\right\}=\left\{1,2,3\right\}$. The choice
of $M=3$ is consistent with the dimensional reduction hypothesis of Bonabeau et. al.\ \cite{bonabeau99} 
which states that clustering is more likely at low effective dimensions. We also mentioned that 
interesting clustering behavior has been observed in real-world systems wherein three different
species of individuals are mixed within the same territory \cite{bonabeau95,bonabeau99,altmann70}.

Fig.\ \ref{fig::cluster-config} presents typical configurations of a cluster system ($M=3$) 
at different $\alpha$ values.
A relatively high local density of agents is found in the lattice interior due to the ``free" boundary
condition which eliminates all agents that wander beyond the lattice boundaries. We measure the cluster
size $s$ (in cell units) using the Hoshen-Kopelman algorithm \cite{hoshen76}. A histogram 
($\textrm{bin size}=1$) of the average cluster sizes is calculated for every $\alpha$-value using 
10 trials. The cluster-size distribution is fitted by a power-law function over a finite range of
sizes, $s_{min}\leq s\leq s_{max}$, where $s_{min}$ and $s_{max}$ are the minimum and maximum cluster
size, respectively. We determine the $\tau$ value from the best fit curve and use it to characterize
the cluster-size distribution. Table \ref{tab::model} presents the relation between the allelomimesis measure
$\alpha$ with exponent $\tau$.

We also determine the average duration $\left<Q\right>$ that an agent remains in a particular state
and correlate it with the cluster size to which the agent belongs. $\left<Q(s)\right>$ is the amount
of time that an agent stays in a cluster of size $s$. It can serve as a rough measure of cluster
stability.

\subsection{Data from Real-World Cluster Systems}
The predictions of our model are compared with measurements taken from different real-world scale-free
cluster systems (see Table \ref{tab::real-world}). In our model, group dynamics is confined within
a two-dimensional plane which is applicable to real-world clusters formed by human beings and 
terrestrial animals. Our model remains valid even to fishes which have been found not to utilize
fully the three-dimensional character of oceanic space \cite{bonabeau99}.

Cluster-size distributions are normally taken from direct-count observations and presented as absolute
frequency $D\left(s\right)$ plots when the number of data points is sufficiently large, or as
cumulative-frequency plots when the data set is sparse. To within a pre-defined accuracy, both methods
yield the same $\tau$ value of the $D\left(s\right)$ plot. Cluster-size data from different kinds of
animals are presented commonly as direct-count values and data reliability depends heavily on the 
accuracy of spotting the correct number of members in an animal group. Data on free-swimming tuna,
buffalo, and clupeid fish are taken from Bonabeau's study \cite{bonabeau95,bonabeau99} while those of
Serengeti lions and cheetahs were sourced from Schaller \cite{schaller72}. Information on baboons is
accrued from several separate studies by Altmann and Altmann \cite{altmann70} on three different baboon species.
Marmot data were quoted from Grimm et. al.\ \cite{grimm03}. Cluster data about dolphins and manatee were
obtained from direct-counting and their reliability was limited by visibility.

Information about city populations and slum areas were taken from Brinkhoff \cite{brinkhoff} and 
Sobreira and Gomes \cite{sobreira01}, respectively. The available data sets were plotted as a 
cumulative frequency distribution, $C\left(\geq s\right)=\sum_{s'}N\left(s'\right)$, where 
$N\left(s'\right)$ are the number of clusters of size $s'$. The summation is taken from $s'=s$ to 
the largest available cluster size $s_{max}$. Exponent $\tau$ is calculated using the property of
power-law distributions which states  that if $\tau '$ is the exponent of $C\left(\geq s\right)$ then it
follows that $\tau=\tau'+1$ \cite{burroughs01}. Data about the employee sizes of U.S. firms were taken
from Axtell \cite{axtell01} while those from Swedish, Japanese and British firms were obtained from
Johansson \cite{johansson97} and Simon and Bonini \cite{simon58}, respectively.

We also studied cluster size data of inanimate systems ($\alpha\approx 0$) such as gene families in
various kinds of bacteria \cite{huynen98}, olivine crystal sizes in xenoliths of the lithospheric
mantle \cite{armienti02}, and galaxy clusters \cite{jarrett}.

\section{Experimental Results}
\label{sec::results}

\begin{figure} 
 \includegraphics[width=8.6cm]{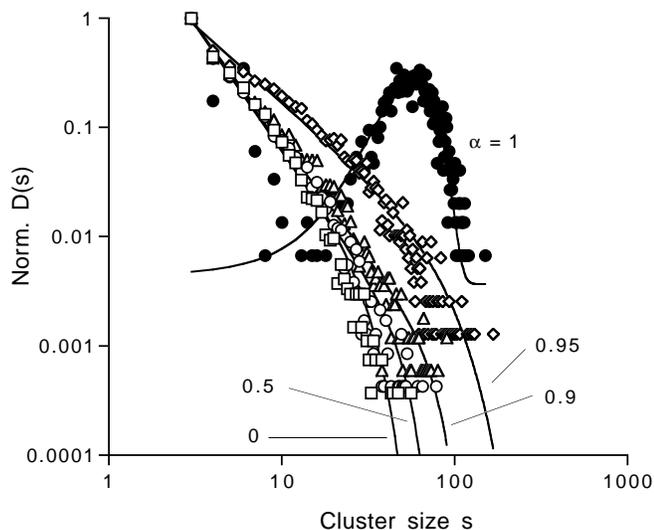}
 \caption{Agent-based modeling ($N=100, M=3$). $D\left(s\right)$  plots for different $\alpha$
 values. Solid lines represent data fits with Fischer scaling function.
 \label{fig::csd}}
 \end{figure}

Fig.\ \ref{fig::csd} shows plots of normalized cluster-size frequency distributions, 
$D\left(s\right)=N\left(s\right)/N\left(s_{min}\right)$, for different $\alpha$ values
($0\leq\alpha\leq0.996$). Simplex projection curve fits are derived (solid lines) using the Fischer
scaling function \cite{bauer88}, $F\left(s\right)=A s^{-\tau}\exp\left(-bs_c\right)$, where $s_c$ is
the cutoff size, $A$ and $b$ are constants and $s_{min}=3\leq s\leq s_c$. Sizes $s=1$ and $s=2$ are
excluded from the curve fitting procedure to minimize deviations from the power-law distribution
caused by boundary effects and finite agent population. Note that the $D\left(s\right)$ plot obtained
with $\alpha =1$ is Gaussian-like with a characteristic size at $s\approx 56$.

\begin{figure} 
 \includegraphics[width=8.6cm]{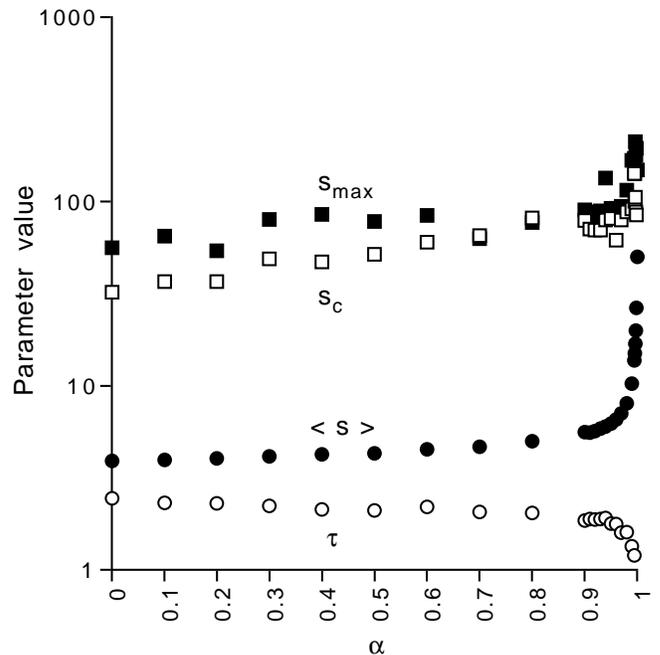}
 \caption{Agent-based modeling, exponent $\tau$ (circles), 
 mean cluster size $\left<s\right>$ (filled circles),
 cutoff-size $s_c$ (squares) and maximum
 cluster size $s_{max}$ (filled squares) versus the allelomimesis measure $\alpha$.
 \label{fig::vs-alpha}}
 \end{figure}

Fig.\ \ref{fig::vs-alpha} plots the dependence of $\tau$ with $\alpha$ (circles) which shows that $\tau$ is 
independent of $\alpha$ for $\alpha<0.9$. However, as $\alpha\rightarrow 1$, $\tau$ rapidly decreases
to zero indicating a nonlinear relation between $\tau$ and $\alpha$. Also shown is the dependence of
the mean cluster size $\left<s\right>=M_1/M_0$, with $\alpha$ (filled circles) where $M_0$ and $M_1$
are the zeroth- and first-order moments of the size distribution $N\left(s\right)$, respectively.

The sharp variation of $\tau$ and $M_1/M_0$ as $\alpha\rightarrow 1$ implies a rapid increase in
clustering among agents, hence, a greater probability of the formation of large clusters.
Fig.\ \ref{fig::vs-alpha} also plots $s_c$ (squares) and $s_{max}$ (filled squares) as a function 
of $\alpha$ that also reveal rapid
and nonlinear increases for both parameters as $\alpha\rightarrow 1$. The plot behavior of $\tau$, 
$M_1/M_0$, $s_c$ and $s_{max}$ against $\alpha$ consistently indicate that strongly-allelomimetic
agents are capable of forming large, compact and considerably stable associations.

\begin{figure} 
 \includegraphics[width=8.6cm]{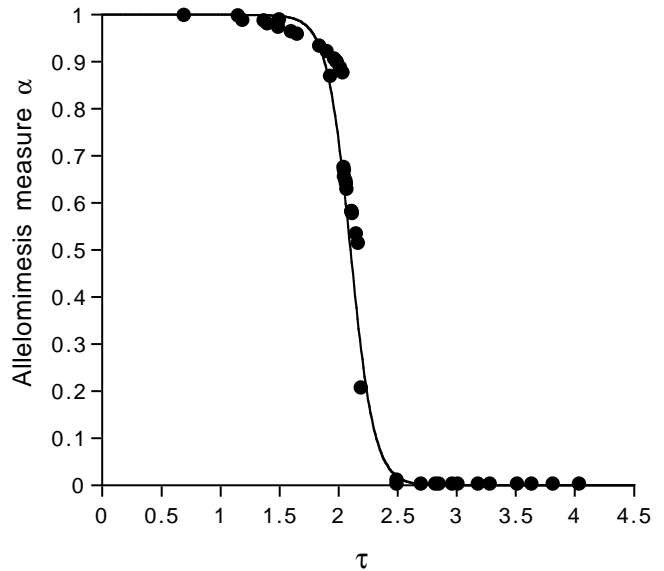}
 \caption{Comparison of numerical prediction (solid $\alpha\left(\tau\right)$ curve) with data
 (circles) from 32 real-world cluster systems. Measured $\tau$ values are plotted according to
 listing order of Table \ref{tab::real-world}.
 \label{fig::alpha-vs-tau}}
 \end{figure}

Fig.\ \ref{fig::alpha-vs-tau} plots the pair of $\alpha$-$\tau$ values that were generated using our agent-based model.
The nonlinear $\alpha\left(\tau\right)$ curve strongly approximates a Fermi distribution,
\begin{equation}
\label{eq::fermi}
\alpha\left(\tau\right)=\gamma\left\{1+\exp\left[\beta\left(\tau-\tau_c\right)\right]\right\}^{-1}
\end{equation}

\noindent with $\gamma\approx 1$, $\beta = 0.104$ and $\tau_c = 2.15$. Also plotted in 
Fig.\ \ref{fig::alpha-vs-tau} are the 
measured $\tau$ values (filled circles) from thirty-two selected real-world cluster systems in the
order presented in Table \ref{tab::real-world}.

The corresponding $\alpha$ is interpolated from the given measure $\tau$ value of a real-world cluster
system via Eq.\ \ref{eq::fermi} (reduced $\chi^2 = 0.00529$; $R^2=0.9565$). The behavior of Eq.\ \ref{eq::fermi}
hints to the presence of three general types of allelomimetic interactions, 1) \emph{Blind allelomimesis}
($\alpha\approx1$) where agents are most likely to copy conspecifics, 2) \emph{Information-use allelomimesis}
($\alpha\propto\tau$) where agents are deliberate in their decisions to copy conspecifics, and 3)
\emph{Non-allelomimetic} ($\alpha\approx0$) where agents do not possess the social attribute to copy
their neighbors.

Our findings are consistent with the experiment-based classifications proposed earlier by Wagner and
Danchin \cite{wagner03}. Information-use copying may be considered an advance (evolutionary) trait found
in humans. It plays a critical role in the expansion of business firms and cities \cite{axtell01,makse95}.
Interestingly, our model indicates that the growth of slum areas is driven by blind copying rather than
deliberate decisions based on available information.

Many animal clusters (e.g., fish schools, buffalos) benefit from blind copying ($0.98<\alpha<0.99$).
Herding is more likely to emerge quickly in systems with strongly-allelomimetic agents. On the other
hand, baboons have a relatively low $\alpha$ at $0.78$ because they live in heirarchical societies
\cite{altmann70} where higher-ranked individuals are more likely to succeed in imposing their will on
others below them -- allelomimesis is biased towards dominance. The West Indian manatee have a relatively
low $\alpha$ of $0.40$ because they are solitary animals \cite{kadel91} and are unlikely to encounter 
other individuals of the same kind within their lifetime. Cheetahs and lions both have low $\alpha$
of $0.038$ because they are nomadic and prefer to hunt alone or in small packs \cite{schaller72}.
In these animals, the chances of being influenced by their neighbors are rather low.

\begin{figure} 
 \includegraphics[width=8.6cm]{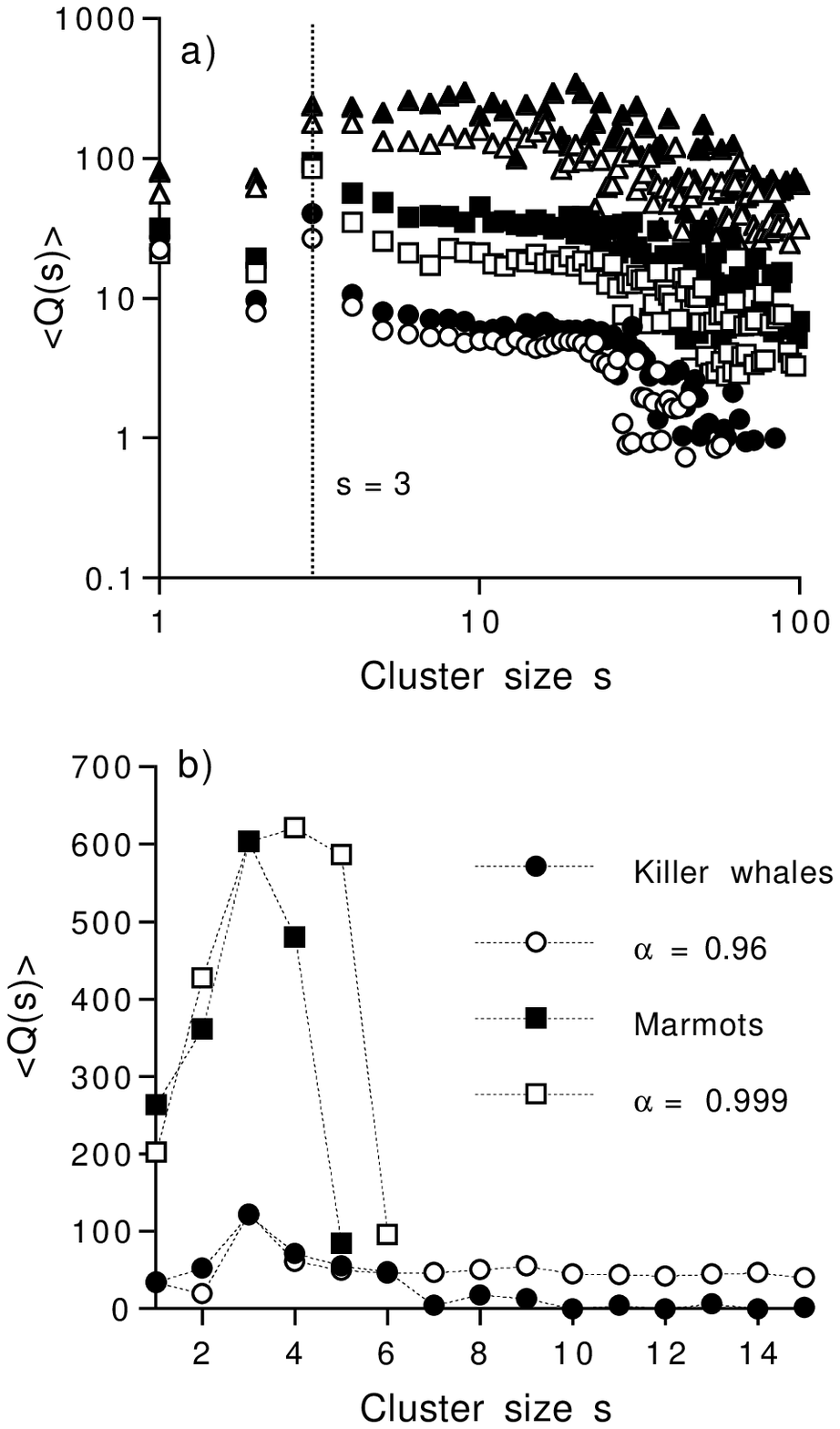}
 \caption{Comparison of numerical predictions with real-world data: a) $\left<Q\left(s\right)\right>$
 from agent-based model at $\alpha=0$ (circles), $0.5$ (filled circles), $0.9$ (squares), $0.95$
 (filled squares), $0.99$ (triangles) and $0.995$ (filled triangles); 
 b) $\left<Q\left(s\right)\right>$ from killer whales (filled circles) and marmots (filled squares), and 
 corresponding $\left<Q\left(s\right)\right>$ from agent-based model at $\alpha=0.96$ (circles) and
 $0.99$ (squares).
 \label{fig::stability}}
 \end{figure}

Fig.\ \ref{fig::stability}a plots the average time duration $\left<Q\right>$ that an agent maintains its state as
a function of cluster size $s$. The $\left<Q\right>$-plots indicate that $\left<Q\right>$ is highest
for clusters that consists of three members ($s=3$) which has been observed in killer whales
(\emph{Orcinus orca}) if we interpret $Q$ as the number of hours of observation in direct count
studies \cite{baird96}. Fig.\ \ref{fig::stability}b presents the dependence of $\left<Q\right>$ with $s$ for killer
whales. Also shown are the $\left<Q\right>$ values predicted by our model for $\alpha=0.96$. The
cluster-size distribution data are unavailable for killer whales and they are assumed to be similar
to dolphins where group size data are available. This assumption is justified because the killer
whale is a member of the dolphin family \cite{culik}. Dolphins exhibit cluster size distributions
($\tau=1.79\pm0.05$) that corresponds to $\alpha\approx0.96$ (see Table \ref{tab::real-world}).

Yellow-bellied marmots (\emph{Marmota flaviventris}) also exhibit optimum cluster stability at
$s=3$ \cite{armitage00}. Fig.\ \ref{fig::stability}b also plots $\left<Q\right>$ versus $s$ where $Q$ is the net
reproductive rate which is directly related to the amount of time spent by individuals as a
group. Also shown are the predicted $\left<Q\right>$ values for $\alpha=0.99$ obtained with a
larger lattice ($N=500$) which is utilized because marmots operate in relatively wide territories
such as steppes, alpine meadows, pastures, or fields \cite{smith}.
 
 \begin{table} 
 \caption{Some $\alpha$ values and resulting cluster-size distribution fitted by a power-law
 with exponent $-\tau$ taken over cluster size-range, $s_{min}\leq s\leq s_{max}$
 \label{tab::model}}
 \begin{ruledtabular}
 \begin{tabular}{lccc}
 $\alpha$ & $\tau$ & $s_{min}$ & $s_{max}$ \\\hline
 $0.0$ & $2.451\pm0.092$ & $3$ & $21$ \\ 
 $0.1$ & $2.318\pm0.096$ & $3$ & $22$ \\ 
 $0.2$ & $2.298\pm0.080$ & $3$ & $22$ \\ 
 $0.3$ & $2.233\pm0.077$ & $3$ & $20$ \\ 
 $0.4$ & $2.127\pm0.085$ & $3$ & $20$ \\ 
 $0.5$ & $2.110\pm0.039$ & $3$ & $22$ \\ 
 $0.6$ & $2.200\pm0.047$ & $3$ & $24$ \\ 
 $0.7$ & $2.060\pm0.072$ & $3$ & $24$ \\ 
 $0.8$ & $2.040\pm0.033$ & $3$ & $27$ \\ 
 $0.9$ & $1.858\pm0.030$ & $3$ & $25$ \\ 
 $0.91$ & $1.887\pm0.048$ & $3$ & $25$ \\ 
 $0.95$ & $1.789\pm0.047$ & $3$ & $27$ \\ 
 $0.99$ & $1.343\pm0.018$ & $3$ & $34$ \\ 
 $0.995$ & $1.199\pm0.015$ & $3$ & $43$ \\ 
 $0.9996$ & $1.045\pm0.016$ & $3$ & $42$ \\ 
 \end{tabular}
 \end{ruledtabular}
 \end{table}

\section{Discussion}
We have introduced a model in which interactions between agents are driven by their likelihood
to imitate one another. The model supports the formation of spatial clusters that obey the 
power-law cluster-size frequency distribution over a wide range of possible $\tau$ values. 
The rules governing allelomimetic interactions are few and simple. 
A single parameter $\alpha$ ($0\leq\alpha\leq 1$) is needed to tune the exponent $\tau$ value over a wide range of values.  
Our model is generic and could explain the broad spectrum of $\tau$ values observed with 
different kinds of real-world scale-free cluster systems (see Table \ref{tab::real-world}).  

A nonlinear relation that is described by a Fermi distribution, exists between the degree of
allelomimetic behavior (as measured by $\alpha$) and $\tau$ which describes the relative 
abundance of the cluster sizes in the system. The $\alpha(\tau)$ curve in Fig.\ \ref{fig::alpha-vs-tau}, 
accurately
tracks the $\tau$ values that have been measured with real-world clusters. Successful correlation
between theory and experimental evidence has been achieved even with a simplified agent-based
model that neglects possible differences in the allelomimetic behavior among agents. We have
assumed that all agents in a given system have the same $\alpha$ and are confined to interact
on a two-dimensional plane.

The nonlinear character of the $\alpha(\tau)$ curve indicates the presence of three general
classes of allelomimetic interactions namely, \emph{blind copying}, \emph{information-use copying},
and \emph{non-allelomimetic}. Our findings are consistent with previous claims that were derived
directly from experimental evidence \cite{wagner03}. Different cluster systems that benefit from
blind copying are characterized by a small $\alpha$-range that is near unity. However, such
systems yield cluster-size frequency distributions with a wide range of possible $\tau$-values
($0\leq\tau< 2$) implying that the relative abundance of the cluster sizes in systems with
strongly-allelomimetic agents is sensitive to small variations in $\alpha$. A similar sensitivity
characteristic also occurs with non-allelomimetic cluster systems such as colloids and
galaxies ($\alpha\approx 0$).

Cluster systems that are formed by humans such as slums, cities, and business firms are
associated with a wide range of $\alpha$ values ($0.1\leq\alpha\leq 0.9$). However, their
$D\left(s\right)$ plots are restricted within a limited spectrum of $\tau$-values ($1.4\leq\tau< 2.16$).
Human beings have developed (by evolution and learning from past mistakes) the capability to
decide on their own or as a collective based on a set of often contending factors. The formation
of slum areas which are anti-social and often illegal entities, is driven by collective action
of informal settlers which is a strongly allelomimetic behavior. On the other hand, business
firms and the cities in Germany and Japan are formed deliberately based on information-driven
plans and project studies. The cluster-size distribution of slum areas tend to be uniform while
those of business firms are likely to be biased towards the small and medium sized (in terms
of employee number).

In cluster formation that arises from information-use copying, the cluster-size frequency 
distribution is weakly sensitive to slight $\alpha$ variations ($\alpha\propto\tau$) unlike
with the other two classes of allelomimetic behavior. The information-based cluster systems
are quite robust -- significant shift in the allelomimetic tendency of the component agents
only results in a slight change of the cluster-size frequency distribution.

Our model has also predicted that with strongly-allelomimetic agents, the most stable cluster
is one with three members ($s=3$). Experimental evidence for this interesting finding has been
found in killer whales and marmots. We think that three is a stable company because it is
the smallest (hence the most economical to maintain) cluster size where the concept of 
majority-based decision remains meaningful.

\section{Conclusions}
Allelomimesis (or its absence) is a generic interaction mechanism between adaptive agents
that could accurately explain the richness of $\tau$ values that has been observed in real-world
scale-free cluster systems. This is possible because allelomimetic interactions between agents
can be described by few and simple local rules.

We have generated an $\alpha(\tau)$ curve that rationalizes the broad spectrum of observed
$\tau$ values. The curve may be utilized to formulate effective strategies in wildlife 
conservation, urban planning, and even product marketing. Allelomimesis-based interaction
could also explain the existence of a preferred cluster size of $s=3$ in strongly-allelomimetic
animals such as killer whales and marmots.

In the real world, it is not unusual to find several cluster systems occupying a common 
habitat. That each of them can be analyzed with a single model is proof of the underlying
interconnectivity of animate and inanimate clusters. The availability of a \emph{universal} 
mechanism for adaptation is vital in the formulation of effective strategies in wildlife
preservation, environmental management, urban planning, economics, and even politics.

Allelomimesis in endangered species may be enhanced to favor large-cluster formation where
reproductive success is greater since survivorship is directly related to group size
\cite{parrish99}. Urban overcrowding may be reduced with initiatives that discourage
allelomimesis in humans. An efficient army or a successful beauty product may be developed
by strategies that favor blind obedience and mass mimicry, respectively.

\bibliography{xxx_allelomimesis}
\end{document}